\documentclass[twocolumn,showpacs,prl,amsmath,amssymb,floatfix]{revtex4}

\usepackage{dcolumn}
\usepackage{amsmath,amssymb}
\usepackage{bm}
\usepackage{chngpage}
\usepackage{epsfig}
\usepackage{slashed}
\usepackage{dsfont}

\begin{document}
\title{Perturbative analysis of the conductivity in disordered mono-- and bilayer graphene}
\author{Andreas Sinner and Klaus Ziegler}
\affiliation{Institut f\"ur Physik, Universit\"at Augsburg}
\date{\today}

\begin{abstract}
The DC conductivity of monolayer and bilayer graphene is studied perturbatively for different types of disorder. In the case of monolayer, an exact cancellation of logarithmic divergences occurs for all disorder types. The total conductivity correction for a random vector potential is zero, while for a random scalar potential and a random gap it acquires finite corrections. We identify the diagrams which are responsible for these corrections and extrapolate
the finite contributions to higher orders which gives us general expressions for the conductivity of weakly disordered monolayer graphene. In the case of bilayer graphene, a cancellation of all contributions for all types of disorder takes place. Thus, the minimal conductivity of bilayer graphene turns out to be very robust against disorder.

\end{abstract}
\maketitle

{\it Introduction.} Monolayer graphene (MLG) represents a monoatomic sheet of carbon atoms arranged in a honeycomb lattice with lattice spacing $a$ and next--neighbor hopping energy $t\approx 2.8$~eV. The transport properties of charge--neutral MLG are characterized by the semimetallic behavior with a point--like Fermi surface at two nodes (valleys) and linear low--energy dispersion in the vicinity of these 
valleys. This remarkable fact is one reason for the outstanding electronic properties of ML 
graphene~[\onlinecite{Novoselov2005,Novoselov2006,Katsnelson2006}]. Perhaps the most prominent transport property 
of ML graphene is the minimal conductivity $\bar\sigma^{}_0=e^2/h\pi$ exactly at the Dirac point which has been observed in a number of
experiments~[\onlinecite{Novoselov2005,Zhang2005,Novoselov2006}]. Bilayer graphene (BLG) represents two ML honeycomb lattices with  Bernal stacking, where the interlayer hopping processes are allowed with the energy $t^{}_\bot\approx 0.4$~eV. The main difference between MLG and 
BLG is that the low-energy excitations of the latter have a quadratic spectrum in the vicinity of the 
valleys~[\onlinecite{Ohta2006}]. This difference causes a factor of two for the DC conductivity $\bar\sigma^{}_0 = 2e^2/h\pi$. Experimentally, both values seem to depend only 
very weakly on disorder or thermal fluctuations~[\onlinecite{Ohta2006,Geim2007,Tan2007,Chen2008,Morozov2008}], which is supported by  field-theoretical studies for various types of disorder~[\onlinecite{Fradkin1986,Ludwig1994,Ziegler2008,Ziegler1997,Ziegler2009,Sinner2011,Giuliani2011}].

To the best of our knowledge, a systematic diagrammatic analysis of the conductivity of disordered graphene has not
been performed so far. Usually, only certain types of diagrams are taken into account. Such approximations cannot be considered as fully controllable, since each diagram in the perturbative expansion exhibits logarithmic divergences. Therefore, the final result of calculations must crucially depend on a correct counting of diagrams. It is the purpose of this paper to demonstrate this for the conductivity calculated within the Kubo formalism. 

\begin{figure}[t]
\includegraphics[width=6cm]{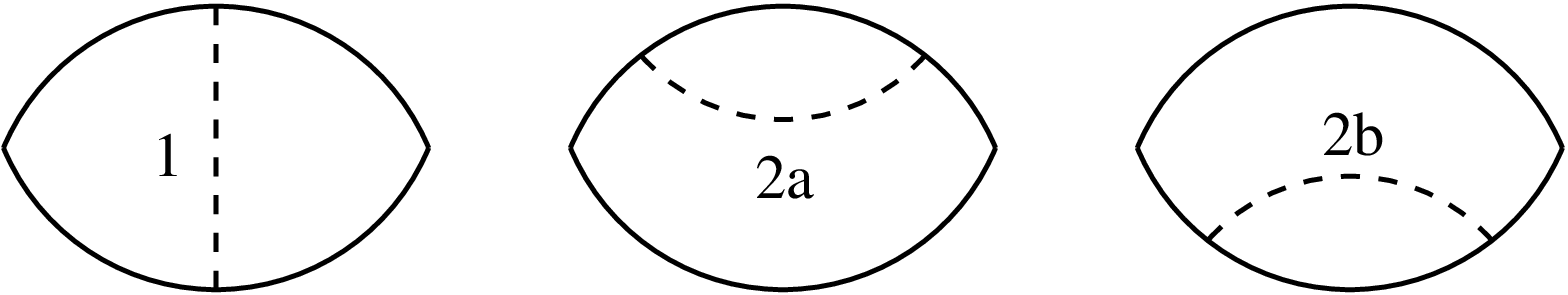}  
\caption{Two--loop corrections of the two--particle Green's function. Each diagram appears twice in the expansion.
Solid lines in the upper/lower bow correspond to Green's functions of clean graphene ($\hat{u}=0$)
$G^{}(-z)/G^{}(+z)$ and slashed lines to the disorder.} 
\label{fig:Series1stOrd}
\includegraphics[width=6cm]{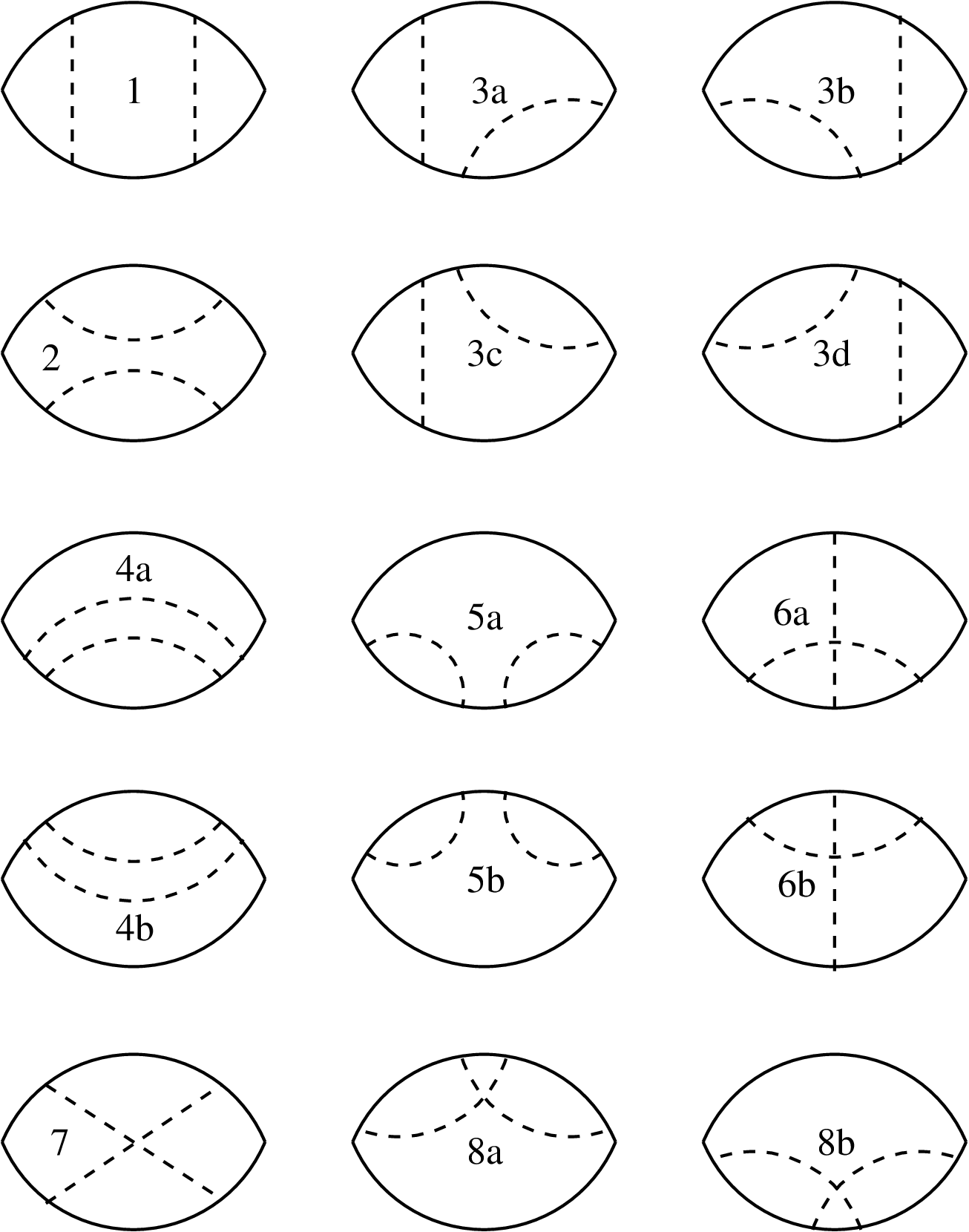}  
\caption{Three--loop corrections of the two--particle Green's function. Each diagram appears four times. 
Generally, the combinatorial factor of each $n$th order diagram is $2^n$.} 
\label{fig:Series2ndOrd}
\end{figure}

{\it Kubo formula.} Within the linear response theory, the conductivity of graphene per a spin and valley projection can be 
approximated for low frequencies ($\omega\sim0$) by the Kubo formula \cite{Abergel2010}

\begin{eqnarray}
\label{eq:Cond1}
\bar\sigma(\omega) = -\omega^2\frac{e^2}{2h}C_g(\omega),
\end{eqnarray}
where
\begin{equation}
\label{eq:Displacement}
C_g(\omega)=
\sum_{r} r^2_k { \rm Tr} 
\left\langle G^{}_{r0}\left(i\epsilon+\frac{\omega}{2}\right)G^{}_{0 r}\left(-i\epsilon-\frac{\omega}{2}\right)\right\rangle_g
\hspace{5mm}
\end{equation}
is the analog of the (mean square) displacement function of a classical random walk. 
The spin--valley degeneracy should be taken into account by multiplying it with an additional factor 4. The symbol $r^{}_k$ denotes the $k$ component of the position operator; the brackets $\langle ...\rangle_g$ mean average with respect to disorder of strength $g$.
The expression $G^{}_{rr^\prime}(z)G^{}_{r^\prime r}(-z)$ is referred to as the two--particle Green's function and is commonly depicted diagrammatically as a closed loop 
\setlength{\unitlength}{1cm}
\begin{picture}(0.7, 0.3)
  \qbezier(0.6, 0.1)(0.3,0.4)(0, 0.1) 
  \qbezier(0.6, 0.1)(0.3,-0.2)(0, 0.1)
\end{picture}.
Here, $G$ denotes the one--particle Green's function with 
\begin{eqnarray}
\label{eq:GF}
G^{-1}(z) = \hbar z + \sigma\cdot\nabla + v^2\hat{u}_r,
\end{eqnarray}
where the vector $\sigma$ consists of Pauli matrices $\sigma_{1,2}$ and $\hat{u}^{}_r$ represents a disorder potential. The operator of kinetic energy reads $\nabla^{}_i=i\hbar v\partial_i$ for MLG with the Fermi velocity $v=\sqrt{3}ta/2\hbar$, and $\nabla_1=\hbar^2(\partial^2_1-\partial^2_2)/2\mu$ and $\nabla_2 = \hbar^2 \partial_1\partial_2/\mu$ for BLG, where $\mu=2t_\bot\hbar^2/3t^2 a^2$ denotes the electron band mass. 
For BLG we also define a characteristic velocity $v=\hbar/a\mu=3t^2a/2t_\bot\hbar$ which appears in the Green's function. 
Here, we consider three disorder types: 1) random scalar potential, $\hat{u}_r=V_r\sigma^{}_0$; 2) random vector potential
$\hat{u}^{}_r=\sigma^{}\cdot A^{}_r$ and 3) random gap $\hat{u}_r = \delta M_r\sigma^{}_3$. We assume disorder
to be Gaussian correlated with zero mean, i.~e. $\langle \hat{u}^{}_r\rangle=0$, $\langle
\hat{u}^{}_r\hat{u}^{}_{r^\prime}\rangle = (\hbar/v)^2 (g/mv^2)\delta(r-r^\prime)$ ($\langle
\hat{u}_r\hat{u}_{r^\prime}\rangle = (\hbar/v)^2 (g/m v^2)\delta_{ij}\delta(r-r^\prime)$ for random vector
potential), with corresponding $v$ and $m=m^{}_e$, the bare electron mass for MLG ($m^{}_ev^2~\approx~6$eV), and $m= \mu$ for BLG ($\mu v^2\approx 30$eV). Our assumption suggests $g/(mv^2)\ll 1$. Below we use a unit system with $\hbar=1$, $e^2/h=1$, $m^{}_e=1$, $2\mu=1$ and $v=1$. 

Before we embark on the perturbative calculation, we briefly discuss the status quo of the field--theoretical approach.
In the case of a random vector potential, representing random ripples in graphene, a bosonized replica approach yields for MLG 
$\bar\sigma = \bar\sigma^{}_0$~[\onlinecite{Ludwig1994}].
The random gap case is of particular interest because it describes a metal--insulator transition due to the opening of the gap $m$.
The displacement function $C_g(m,\omega)$ can be evaluated in this case by replacing the random field
$\delta M_r$ by a more general random field. This mapping enables us to search for saddle 
points in a multidimensional manifold. It turns out that not just a single saddle point exists but a whole 
saddle--point manifold. As a result, we have one massless 
mode \cite{Ziegler2009}. The latter creates an $\omega^{-2}$ singularity in $C_g(m,\omega)$, 
which cancels the $\omega^2$ factor in the conductivity of Eq.~(\ref{eq:Cond1}). There are also massive
modes around the saddle--point manifold. Altogether this leads to the following scaling form: 
\begin{equation}
C_g(m,\omega)=\frac{(\omega+2i\eta)^2}{\omega^2} C_0(m/2,\omega+2i\eta)K_g,
\end{equation}
with the displacement function of the pure system
\begin{equation}
\label{eq:DisplFunc}
C_0(m,\omega)=\frac{2}{\pi}\frac{1}{4m^2-\omega^2} \ .
\end{equation}
$K_g$ is the contribution of the massive modes. Although its form is unknown for $g>0$, it is always finite with 
$K_0=1$ for $g=0$. The scattering rate 
$\eta$ is given by 
$\eta=(m^2-m_c^2)\Theta(m_c^2-m^2)/4$ with $m_c=\Lambda e^{-\pi/g}$ \cite{Ziegler2009}.
It should be noticed that the $\omega^{-2}$ singularity of $C_g$ disappears for both MLG and BLG for $\eta=0$ and $m>0$.
In terms of the conductivity of Eq.~(\ref{eq:Cond1}) this yields for $\omega\sim0$ eventually
\begin{equation}
\label{eq:SigNonPert}
\bar\sigma\sim {\bar\sigma}^{}_0 K_g\left(1-\frac{m^2}{m_c^2}\right)\Theta(m_c^2-m^2)
 \ .
\end{equation}
This result indicates that $\bar\sigma$ vanishes for any $m>0$ in perturbation theory because
$m_c$ is always zero in the latter.
On the other hand, for $m=0$ the perturbation theory should reproduce $\bar\sigma =\bar\sigma^{}_0 K_g$, since the 
$m_c$ drops out of the conductivity in this case.

{\it Perturbative theory:} Now we evaluate the displacement function perturbatively in powers of the disorder 
strength $g/2$. In order to perform these calculations, the translational invariance broken by disorder has to be restored by averaging over the Gaussian ensemble. Here, we perform this averaging using a diagrammatic representation. 
The first order conductivity corrections arise from the graphs depicted in Fig.~\ref{fig:Series1stOrd}. 
Results of the evaluation of these diagrams are shown in Table~\ref{tab:Corrections1stOrd}, where we only retain constant contributions and logarithmically divergent terms. The second--order (three--loop order) diagrams contributing to the conductivity are shown in Fig.~\ref{fig:Series2ndOrd} and results of their evaluation are summarized in Table~\ref{tab:Corrections2ndOrd}. Technical details of the evaluation can be found in the supplementary material~[\onlinecite{Supplementary}]. The total combinatorial factor for each topological class of diagrams is $2^n$. Each topological class of diagrams exhibits in turn further degeneracy due to diagram symmetries: The degeneracy factor of the first--order topological class 1 (Fig.~\ref{fig:Series1stOrd}) and of the second--order topological classes 1, 2 and 7 (Fig.~\ref{fig:Series2ndOrd}) is one, while that of the first--order topological 
class 2 (Fig.~\ref{fig:Series1stOrd}) and of the second--order topological classes 4, 5, 6 and 8 (Fig.~\ref{fig:Series2ndOrd}) 
is two and degeneracy factor of the second--order topological class 3 is four.  All these factors
should be carefully taken into account. 

First we discuss results for monolayer graphene. To the two--loop order, contributions to the displacement function  from all diagrams reveal divergences $\sim\omega^{-2}\ln\Lambda/\omega$. Here, $\Lambda$ denotes the UV--cutoff and $\omega$ the imaginary frequency in contrast to Eqs.~(\ref{eq:Cond1}) and~(\ref{eq:Displacement}). The emergence of the logarithms is due to the divergence of the loop integrals 
\begin{equation}
\label{eq:Divergence} 
I = 
\int\frac{d^2k}{(2\pi)^2}~G^{}_k(z) \sim \ln\frac{\Lambda}{\omega},
\end{equation}
while the singularity $\omega^{-2}$ appears in the displacement function due to re--scaling of the factor $r^2_k$ in Eq.~(\ref{eq:Displacement}) ~[\onlinecite{Supplementary}], 
\begin{widetext}
\begin{center}
\begin{table}[t]
\begin{tabular}{cccc}
     Diagram class \,\,\,\,   &  {\rm Scalar disorder} &  {\rm Gap disorder}  &   {\rm Vector disorder}   \\
\\
({\rm \ref{fig:Series1stOrd}}, 1)  $\times$ 1
&  
\hspace{0.5cm}
$
\alpha[1+2\ell]/2
\hspace{5mm}
(2\beta)
$
\hspace{0.5cm}
& 
\hspace{0.5cm}
$
-\alpha[1-2\ell]/2
\hspace{5mm}
(2\beta)
$
\hspace{0.5cm}
&
\hspace{0.5cm}
$\displaystyle 2\alpha\ell
\hspace{5mm}
(4\beta) 
$  
\hspace{0.5cm}
\\
\\
({\rm \ref{fig:Series1stOrd}}, 2) $\times$ 2
& 
$\hspace{3mm}
-\alpha\ell
\hspace{11mm}
(-2\beta)
$
&
$\hspace{6mm}
-\alpha\ell
\hspace{11mm}
(-2\beta)
$ 
& 
$\displaystyle -2\alpha \ell 
\hspace{4mm}
(-4\beta)
$ \\
\\
{\rm Total} $\times$ 2
& 
$\hspace{4mm}
\alpha
\hspace{16mm}
(0)
$  
&
$\hspace{5mm}
-\alpha
\hspace{14mm}
(0)
$  
&
$\;\;\; 0
\hspace{8mm}
(0)
$
\\
\end{tabular}
\caption{Conductivity corrections from the first--order classes of diagrams (Fig.~\ref{fig:Series1stOrd}) for MLG (BLG) in corresponding $\bar\sigma^{}_0$  units~[\onlinecite{Supplementary}]. MLG results are shown for $\Lambda\gg\omega$. Here we use the shorthands $\alpha=g/2\pi$, $\ell=\log(\Lambda/\omega)$, and $\beta = g/8\omega$. The degeneracy factors for each diagram and combinatorial factor 2 are taken into account (first column).}
\label{tab:Corrections1stOrd} 
\end{table}
\end{center}
\begin{center}
\begin{table}[t]
\begin{tabular}{cccc}
     Diagram class       &  {\rm Scalar disorder} &  {\rm Gap disorder}  &   {\rm Vector disorder}   \\
\\
({\rm \ref{fig:Series2ndOrd}}, 1) $\times$ 1  
&  
$\displaystyle 
\alpha^2[1 + 4\ell + 6\ell^2]/8
\hspace{5mm}
(3\beta^2/2)
$
& 
$\displaystyle
\alpha^2[1 - 4\ell + 6\ell^2]/8 
\hspace{5mm}
(3\beta^2/2)
$
&
$\hspace{6mm}
3\alpha^2\ell^2
\hspace{11mm}
(6\beta^2)
$\\\\
({\rm \ref{fig:Series2ndOrd}}, 2) $\times$ 1 
& 
$\hspace{6mm}
5\alpha^2\ell^2/12
\hspace{12mm}
(5\beta^2/6)
$
&
$\hspace{6mm}
5\alpha^2\ell^2/12
\hspace{12mm}
(5\beta^2/6)
$ 
& 
$\hspace{6mm}
5\alpha^2\ell^2/3
\hspace{8mm}
(10\beta^2/3)
$ 
\\\\
({\rm \ref{fig:Series2ndOrd}}, 3) $\times$ 4
&
$\hspace{1mm}
-\alpha^2\ell[1+\ell]
\hspace{12mm}
(-3\beta^2)
$
&
$\hspace{4mm} 
-\alpha^2\ell^2
\hspace{16mm}
(-3\beta^2)
$
&
$\displaystyle 
-2\alpha^2\ell[1+2\ell]
\hspace{5mm}
(-12\beta^2)
$
\\\\
({\rm \ref{fig:Series2ndOrd}}, 4) $\times$ 2
&
$\hspace{3mm}
\alpha^2\ell[1-\ell]/2
\hspace{10mm}
(2\beta^2/3)
$
&
$\hspace{2mm}
\alpha^2\ell[1-\ell]/2
\hspace{11mm}
(2\beta^2/3)
$
&
$\hspace{5mm}
2\alpha^2\ell[1-\ell]
\hspace{8mm}
(8\beta^2/3)
$
\\\\
({\rm \ref{fig:Series2ndOrd}}, 5) $\times$ 2
&
$\hspace{4mm}
\alpha^2\ell^2/3
\hspace{16mm}
(0)
$
&
$\hspace{4mm}
\alpha^2\ell^2/3
\hspace{16mm}
(0)
$
& 
$\hspace{3mm}
4\alpha^2\ell^2/3
\hspace{13mm}
(0)
$
\\\\
({\rm \ref{fig:Series2ndOrd}}, 6) $\times$ 2
& $\hspace{8mm}0$
& $\hspace{8mm}0$
& $\hspace{8mm}0$
\\\\
({\rm \ref{fig:Series2ndOrd}}, 7) $\times$ 1
& $\hspace{8mm}0$
& $\hspace{8mm}0$
& $\hspace{8mm}0$
\\\\
({\rm \ref{fig:Series2ndOrd}}, 8) $\times$ 2
& $\hspace{8mm}0$
& $\hspace{8mm}0$
& $\hspace{8mm}0$
\\\\
{\rm Total} $\times$ 4
& 
$\hspace{4mm}
\alpha^2/2
\hspace{19mm}
(0)
$  
&
$\hspace{6mm} 
\alpha^2/2
\hspace{18mm}
(0)
$  
&
$\hspace{8mm}0$
\\
\end{tabular}
\caption{Conductivity corrections from the second--order classes of diagrams (Fig.~\ref{fig:Series2ndOrd}) for MLG (BLG) in corresponding $\bar\sigma^{}_0$  units. All shorthands as above. The degeneracy factors and combinatorial factor 4 are taken into account (first column).}
\label{tab:Corrections2ndOrd} 
\end{table}
\end{center}
\end{widetext}

\noindent
but disappears in the conductivity because of the factor $\omega^2$ in Eq.~(\ref{eq:Cond1}), as in the field--theoretical approach discussed above. At the three--loop level, contributions from diagrams containing intercrossing impurity lines (diagrams 6, 7 and 8 in Fig.~\ref{fig:Series2ndOrd}) vanish after angular integration. Contributions to the displacement function arising from each diagram with non--crossing impurity lines diverge $\sim \omega^{-2}(\ln\Lambda/\omega)^2$, with some of them revealing subdominant divergence $\sim\omega^{-2}\ln\Lambda/\omega$.  However, to both orders $g$ and $g^2$, the sum over all conductivity contributions is finite, i.~e. singularities to both orders $\sim\ln\Lambda/\omega$ and $\sim (\ln\Lambda/\omega)^2$ cancel each other exactly. For the random vector potential, the conductivity correction is zero to both two-- and three--loop order. This is in accord with the findings of Refs.~[\onlinecite{Ludwig1994,Sinner2011,Giuliani2011}]. For random gap and random scalar potential to both orders, finite conductivity corrections are generated only by ladder diagrams, i.~e. by diagrams 1 in Figs.~\ref{fig:Series1stOrd} and~\ref{fig:Series2ndOrd}. Provided that the cancellation of singularities holds to higher orders as well, the analysis of higher order ladder diagrams yields the following general expression for the $n$th order ($n\geqslant1$) conductivity correction:
\begin{equation}
\bar\sigma^{}_0\frac{(\pm1)^n}{2^{n-1}}\left(\frac{g}{2\pi}\right)^n,
\end{equation}
with $+$ for the random scalar potential and $-$ for the random gap. This expression has been verified to the fourth order in perturbative expansion (five--loop order).  
The sum over $n\geqslant1$ converges and we obtain for the conductivity
\begin{equation}
\label{eq:Solution}
\bar\sigma^{}_{V,M}=\bar\sigma^{}_0 \frac{\displaystyle 1\pm\frac{g}{4\pi}}{\displaystyle 1\mp\frac{g}{4\pi}} \ ,
\end{equation}
for random scalar potential ($+$) and random gap ($-$), correspondingly. It should be noticed that the conductivity
is enhanced (reduced) by scalar--potential (gap) disorder. This is plausible because the fluctuations of the
former add particles to the system at the Dirac point, whereas the latter opens a fluctuating gap that should
reduce the contribution to the conductivity. 

In the case of bilayer, the integral in Eq.~(\ref{eq:Divergence}) converges for $\Lambda\to\infty$, giving $I\sim{\rm const}$. Therefore, corrections to the displacement function arising from each $n$th order diagram become proportional to $\omega^{-2-n}$. Formally, this leads to the singularity $\sim\omega^{-n}$ of each $n$th order conductivity correction for small $\omega$, as it is shown in Tables~\ref{tab:Corrections1stOrd} and~\ref{tab:Corrections2ndOrd} (values in brackets). Details of the evaluation of the first--order diagrams are summarized in~[\onlinecite{Supplementary}]. However, the sum over all contributions gives a zero for all disorder types in both first and second order of perturbation theory. Provided this cancellation holds to higher orders as well, the minimal conductivity of bilayer graphene turns out to be very robust with respect to all types of disorder. 

{\it Discussion:} The conductivity formula Eq.~(\ref{eq:Solution}) represents the main result of our work. It has the required form $\bar\sigma=\bar\sigma_0 K_g$. An important feature of this solution is its scale invariance. Indeed, while this expression has been obtained for a finite frequency $\omega$ after performing the limit $\Lambda\to\infty$, formally, the same result follows by keeping the cutoff $\Lambda$ finite and performing the limit $\omega\to 0$. Hence, Eq.~(\ref{eq:Solution}) can be regarded as an asymptotically exact solution of the DC  transport problem in disordered ML graphene. 

For the particular case of the random gap disorder in MLG with zero average gap, our results confirm findings of the recent numerical  works~[\onlinecite{Bardarson2010,Medvedeva2010}]. For $g\sim0$ it reproduces the field theoretical result obtained in [\onlinecite{Ziegler1997}]. The robustness of the minimal conductivity for BLG for zero average mass is in a remarkable agreement with the non--perturbative result obtained in~[\onlinecite{Ziegler2009}]. On the other hand, the perturbation theory fails to describe a metal--insulator transition typical for a two--dimensional electron gas~[\onlinecite{Ziegler2009,Medvedeva2010}]. This is because the scattering rate $\eta$ vanishes in perturbation theory, whereas a nonzero $\eta$ is the parameter which controls the metal--insulator transition according to  Eq.~(\ref{eq:SigNonPert}). Hence, the area of applicability of perturbation theory is restricted to the metallic phase. 

Apart from the contribution to the displacement function shown in Eq.~(\ref{eq:Displacement}), there are two further contributions which emerge from the Kubo formula and might become important sufficiently far away from the Dirac point~[\onlinecite{Ziegler2008}]. In terms of one--particle Green's functions they correspond to the product $G(\pm z)G(\pm z)$. However, close to the Dirac point these contributions can be neglected in comparison to Eq.~(\ref{eq:Displacement}). Although the reason for neglecting them is not evident from the point of view of perturbation theory, the argument is provided by the field theory. The latter demonstrates the absence of massless modes at the saddle--point for these contributions~[\onlinecite{Ziegler1997,Ziegler2008}], i.~e. at the Dirac point they are strongly suppressed and do not contribute to the transport.

In conclusion, we have carefully studied perturbative corrections to the conductivity of disordered monolayer and bilayer graphene for different disorder types. Up to three--loop order we managed to show that in the case of ML graphene logarithmic divergences cancel each other exactly, irrespectively of the disorder type. Thus, the conductivity of weakly disordered monolayer graphene is modified by a finite correction. On the other hand, the minimal conductivity of bilayer graphene does not acquire any corrections for any type of disorder, as
we have demonstrated for the first and second order in the perturbative expansion.

{\it Acknowledgements:} We acknowledge financial support by the DFG grant ZI 305/5--1.

\end{document}


\title{
Supplementary material for\\
 "Perturbative analysis of the conductivity in disordered mono-- und bilayer graphene"}
\author{Andreas Sinner and Klaus Ziegler}
\affiliation{Institut f\"ur Physik, Universit\"at Augsburg}
\date{\today}

\maketitle

\section{Evaluation of the conductivity of clean monolayer graphene}

Below we evaluate Kubo conductivity of clean monolayer graphene from
\begin{eqnarray}
\label{eq:SCond1}
\bar\sigma(\omega) = -\omega^2\frac{e^2}{2h}C_g(\omega),
\end{eqnarray}
where 
\begin{equation}
\label{eq:SCond00}
C_g(\omega) =
\sum_{r} r^2_k { \rm Tr} [G_{r0}(z)G_{0 r}(-z)] = -\left.\frac{\partial^2}{\partial p^{2}_k} \sum_r e^{-ipr} 
{\rm Tr}\{G_{r0}(z)G_{0 r}(-z)\}\right|_{p=0},
\end{equation}
with $z=i\epsilon+ \omega/2$.  After performing Fourier transform we obtain
\begin{eqnarray}
\label{eq:Cond2}
C_g(z) = -   
\int \frac{d^2q}{(2\pi)^2}~\left.\frac{\partial^2}{\partial p^2_k}
{\rm Tr}\left\{G^{}_{q}(z)G^{}_{q+p}(-z)\right\}
\right|_{p=0}.
\end{eqnarray}
Now we perform Wick rotation $z\to iz$, (and $\omega\to i\omega$ in the first term of Eq.~(\ref{eq:SCond1})), and obtain, after tracing over the spinor space,
\begin{eqnarray}
C_g(z) = - 2
\int \frac{d^2q}{(2\pi)^2}~ \frac{\partial^2}{\partial p^2_k}\left. 
\frac{z^2 + q\cdot(q + p)}{[z^2+(q + p)^2][z^2+q^2]} 
\right|_{p=0}. 
\end{eqnarray}
The derivatives have to be performed first. We notice that
\begin{equation}
\label{eq:Derivative}
\frac{\partial^2}{\partial p^2_k}\frac{A(p)}{B(p)} = \frac{A^{\prime\prime}(p)}{B(p)}
-2\frac{A^\prime(p)B^\prime(p)}{B^2(p)} + 2\frac{A(p)[B^\prime(p)]^2}{B^3(p)} - \frac{A(p)B^{\prime\prime}(p)}{B^2(p)}, 
\end{equation}
and obtain
\begin{equation}
\frac{\partial^2}{\partial p^2_k}\left. 
\frac{z^2 + q\cdot(q + p)}{[z^2+(q + p)^2][z^2+q^2]}\right|_{p=0} = 
-\frac{2z^2}{(z^2+q^2)^3},
\end{equation}
and correspondingly
\begin{equation}
C^{}_g = \frac{1}{2\pi z^2} = \frac{2}{\pi\omega^2}.
\end{equation}
Inserting last expression into the Kubo formula Eq.~(\ref{eq:SCond1}) yields the minimal conductivity of ML graphene $\bar\sigma^{}_0=1/\pi e^2/h$. Evaluation for BLG goes analogously and yields $\bar\sigma^{}_0=2/\pi e^2/h$.

\section{First--order perturbative conductivity corrections}

\begin{figure}[t]
\includegraphics[width=10cm]{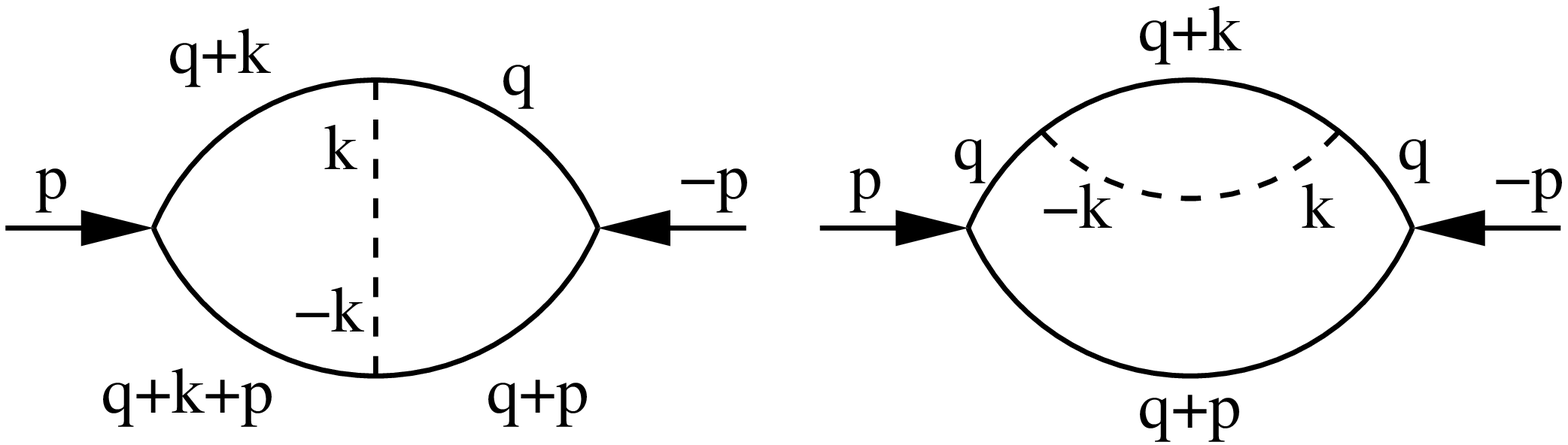}  
\caption{First order diagrams.}
\label{fig:F1}
\end{figure}

Below we calculate conductivity corrections from diagrams depicted in Fig.~\ref{fig:F1}. We start with the case of random gap for MLG. After Wick rotation, first diagram reads 
\begin{widetext}
\begin{eqnarray}
\label{eq:Correction1} 
C^{(1)}_{g,1}(z) = - \frac{g}{2}\left.
\int \frac{d^2q}{(2\pi)^2}\int \frac{d^2k}{(2\pi)^2}~\frac{\partial^2}{\partial p^2_i}
{\rm Tr}\left\{ 
G^{}_{q+k}(-iz)G^{}_{q+k+p}(iz)\sigma^{}_3 G^{}_{q+p}(iz)G^{}_{q}(-iz)\sigma^{}_3
\right\}
\right|_{p=0}, 
\end{eqnarray}
and the second:
\begin{eqnarray}
\label{eq:Correction2} 
C^{(1)}_{g,2}(z) = - \frac{g}{2}\left.
\int \frac{d^2q}{(2\pi)^2}\int \frac{d^2k}{(2\pi)^2}~\frac{\partial^2}{\partial p^2_i}
{\rm Tr}\left\{ 
G^{}_{q+p}(iz)G^{}_{q}(-iz)\sigma^{}_3G^{}_{q+k}(-iz)\sigma^{}_3 G^{}_{q}(-iz)
\right\}
\right|_{p=0},
\end{eqnarray}
\end{widetext}
with Green's functions 
\begin{equation}
G_q(\pm iz) = (\slashed q\pm iz)^{-1} = \frac{\slashed q\mp iz}{z^2+q^2}.
\end{equation}
Here we use the slashed notation $\slashed q=\sigma\cdot q$. We start with Eq.~(\ref{eq:Correction2}). Using the fact that each integration runs from $-\infty$ to $+\infty$ we may shift $k\to k-q$. Then  the $k$--integration acts only on the third Green's function and diverges logarithmically
\begin{equation}
\label{eq:DivPart}
\int \frac{d^2k}{(2\pi)^2}~G^{}_{k}(-iz) = \frac{iz}{4\pi}\ln\left(1+\frac{\Lambda^2}{z^2}\right),
\end{equation}
where the UV-cutoff $\Lambda$ is introduced only temporarily and will be removed at the end of calculation. This quantity couples to the unity matrix and can be pulled out from the trace operator. After performing the trace, the $q$--integral becomes 
\begin{eqnarray}
-\frac{g}{2}\int \frac{d^2q}{(2\pi)^2}~\left.\frac{\partial^2}{\partial p^2_k}\frac{2iz}{(z^2+q^2)^2}\left[1-\frac{p^2}{(z^2+(q+p)^2)} \right]\right|_{p=0} 
= \frac{ig}{4\pi}\frac{\Lambda^2(\Lambda^2+2z^2)}{z^3(\Lambda^2+z^2)^2},
\end{eqnarray}
such that the contribution to the displacement function from this diagram becomes
\begin{equation}
\label{eq:C12}
C^{(1)}_{g,2}(z)=-\frac{g}{16\pi^2}\frac{\Lambda^2(\Lambda^2+2z^2)}{z^2(\Lambda^2+z^2)^2}\ln\left(1+\frac{\Lambda^2}{z^2}\right).
\end{equation}

The evaluation of the expression in Eq.~(\ref{eq:Correction1}) is more cumbersome. First, we can also reshift the momentum $k\to k-q$ as above. Performing some reordering in the numerator we arrive at the following expression: 
\begin{widetext}
\begin{equation}
\label{eq:AppB1}
C^{(1)}_{g,1} = -\left.\frac{g}{2}
\int \frac{d^2q}{(2\pi)^2}\int \frac{d^2k}{(2\pi)^2}~\frac{\partial^2}{\partial p^2_i}
\frac
{\displaystyle {\rm Tr}\left[z^4+z^2\left\{\slashed p\slashed p +\slashed k(\slashed k+\slashed p) +
\slashed q(\slashed q+\slashed p)\right\} + \slashed k(\slashed k + \slashed p)(\slashed q + \slashed p)\slashed q\right]}
{\displaystyle (z^2+k^2)(z^2+q^2)(z^2+(p+q)^2)(z^2+(p+k)^2)}
\right|_{p=0}.
\end{equation}
Using contraction formula for Pauli-matrices
\begin{equation}
\sigma^{}_\alpha\sigma^{}_\beta\sigma^{}_\gamma\sigma^{}_\sigma = \delta^{}_{\alpha\beta}\delta^{}_{\gamma\sigma}-
\delta^{}_{\alpha\gamma}\delta^{}_{\beta\sigma}+\delta^{}_{\alpha\sigma} \delta^{}_{\beta\gamma},
\end{equation}
the trace over the last term in the numerator becomes
\begin{equation}
{\rm Tr}[\slashed k(\slashed k + \slashed p)(\slashed p + \slashed q)\slashed q] = 2(q^2k^2+q^2 k\cdot p + k^2 q\cdot p + p^2 k\cdot q).
\end{equation}
Thus we get from Eq.~(\ref{eq:AppB1})
\begin{equation}
\label{eq:AppB2}
C^{(1)}_{g,1} = -
\left.{g}
\int \frac{d^2k}{(2\pi)^2}\int \frac{d^2q}{(2\pi)^2}~\frac{\partial^2}{\partial p^2_i}
\frac
{\displaystyle z^4+z^2\left[ p^2 + k\cdot( k+ p) +
 q\cdot( q+ p)\right] + q^2k^2+q^2 k\cdot p + k^2 q\cdot p + p^2 k\cdot q}
{\displaystyle (z^2+k^2)(z^2+q^2)(z^2+(p+q)^2)(z^2+(p+k)^2)}
\right|_{p=0}.
\end{equation}
Using Eq.~(\ref{eq:Derivative}) we obtain for each single term 
\begin{equation}
\nn
\begin{array}{ccc}
\displaystyle
\left.\frac{A^{\prime\prime}(p)}{B(p)}\right|_{p=0} = 
\frac{\displaystyle 2(z^2 +  k\cdot q)}
{\displaystyle (z^2 + q^2)^2(z^2 + k^2)^2};
&&
\displaystyle
\left. \frac{A^\prime(p)B^\prime(p)}{B^2(p)} \right|_{p=0} = 
\frac{\displaystyle 2[k^{}_i(z^2+q^2) + q^{}_i(z^2+k^2)]^2}
{\displaystyle (z^2 + k^2)^3(z^2 + q^2)^3}; 
\\\\
\displaystyle
\left.\frac{A(p)[B^\prime(p)]^2}{B^3(p)}\right|_{p=0} = \frac{\displaystyle 4[k^{}_i(z^2+q^2) + q^{}_i(z^2+k^2)]^2}
{\displaystyle (z^2 + k^2)^3(z^2 + q^2)^3}; 
&&
\displaystyle
\left. \frac{A(p)}{B^2(p)}B^{\prime\prime}(p) \right|_{p=0} =
\frac{\displaystyle 2[(z^2+q^2)+4 k^{}_i q^{}_i+(z^2+k^2)]}
{\displaystyle (z^2+k^2)^2(z^2+q^2)^2}.
\end{array}
\end{equation}
Now we are in the position to perform integrals:
\begin{eqnarray}
\nn
-g\int \frac{d^2k}{(2\pi)^2}\int \frac{d^2q}{(2\pi)^2} \left.\frac{A^{\prime\prime}(p)}{B(p)}\right|_{p=0} &=& -\frac{g}{8\pi^2}\frac{\Lambda^4}{z^2(\Lambda^2+z^2)^2};\\
\nn
-g \int \frac{d^2k}{(2\pi)^2}\int \frac{d^2q}{(2\pi)^2}\left. \left[ 
-2  \frac{A^\prime(p)B^\prime(p)}{B^2(p)} + 2 \frac{A(p)[B^\prime(p)]^2 }{B^3(p)}
\right]\right|_{p=0} &=& -\frac{g}{8\pi^2}\frac{\Lambda^4}{z^2(\Lambda^2+z^2)^2}\ln\left(1+\frac{\Lambda^2}{z^2}\right);\\
\nn
-g \int \frac{d^2k}{(2\pi)^2}\int \frac{d^2q}{(2\pi)^2}\left.\left(-\frac{A(p)B^{\prime\prime}(p) }{B^2(p)}\right)\right|_{p=0} &=&  \frac{g}{4\pi^2} \frac{\Lambda^2}{z^2(\Lambda^2+z^2)}\ln\left(1 + \frac{\Lambda^2}{z^2}\right). 
\end{eqnarray}
The contribution to the displacement function from the first--order ladder diagram then reads:
\begin{equation}
\label{eq:C11}
C^{(1)}_{g,1} = 
-\frac{g}{8\pi^2} \frac{\Lambda^4}{z^2(\Lambda^2+z^2)^2} + 
\frac{g}{8\pi^2} \frac{\Lambda^2(\Lambda^2+2z^2)}{z^2(\Lambda^2+z^2)^2}\ln\left(1 + \frac{\Lambda^2}{z^2}\right).
\end{equation}
\end{widetext}
Taking combinatorial factors into account, the total first--order perturbative contribution to the displacement function becomes
\begin{equation}
C^{(1)}_{g}=2C^{(1)}_{g,1}+4C^{(1)}_{g,2} =
-\frac{g}{4\pi^2} \frac{\Lambda^4}{z^2(\Lambda^2+z^2)^2} ,
\end{equation}
where all terms containing logarithms eliminate each other. After removing the cutoff $\Lambda\to\infty$ we obtain a finite conductivity correction 
\begin{equation}
\label{eq:AppBres}
\bar\sigma^{}_1 = \frac{\omega^2}{2}C^{(1)}_g = -\frac{g}{2\pi^2} = -\frac{g\bar\sigma^{}_0}{2\pi}. 
\end{equation}

For the case of random scalar disorder we have to replace Pauli matrices $\sigma^{}_3$ by unity matrices in Eqs.~(\ref{eq:Correction1}) and (\ref{eq:Correction2}). This does not imply any changes on $C^{(1)}_{g,2}$ but changes $C^{(1)}_{g,1}$. At the level of Eqs.~(\ref{eq:AppB1}) and (\ref{eq:AppB2}) this can be seen by the different sign of the term $z^2\slashed p\slashed p$ or $z^2 p^2$, correspondingly, while all other terms remain the same. Then it leads to the change in 
\begin{eqnarray}
\nn
\left.\frac{A^{\prime\prime}(p)}{B(p)}\right|_{p=0} &=& 
\frac{\displaystyle 2(k\cdot q - z^2)}
{\displaystyle (q^2 + z^2)^2(k^2 + z^2)^2},
\end{eqnarray}
with all other terms remaining without any changes. Eventually, we will obtain a finite conductivity correction as in Eq.~(\ref{eq:AppBres}) but with the opposite sign. 

For the case of random vector disorder we have to replace Pauli matrices $\sigma^{}_3$ by $\sigma^{}_\mu$, $\mu=1,2$ in Eqs.~(\ref{eq:Correction1}) and (\ref{eq:Correction2}) and perform an additional summation over $\mu$. Let's evaluate Eq.~(\ref{eq:Correction2}) for this case. Shifting $k\to k-q$ we separate the singular part from the regular one. The singular part remains the same as in Eq.~(\ref{eq:DivPart}). Evaluation of the regular part goes in line with the analogous calculation for the random gap disorder with the only difference that contraction of Pauli matrices $\sigma^{}_\mu$ gives an additional factor 2. Hence, the result for the conductivity correction reads
\begin{equation}
\bar\sigma^{(2)}_1 = - \frac{g}{2\pi^2}\left(1-\frac{\omega^4}{\Lambda^4}\right)\ln\frac{\Lambda}{\omega},
\end{equation} 
where we redefined the cutoff by $\displaystyle \Lambda\to\sqrt{\Lambda^2-\omega^2}/2$.
For the evaluation of Eq.~(\ref{eq:Correction1}) for the case of vector disorder one has to notice that 
\begin{eqnarray}
\sum_\mu\sigma^{}_\mu \slashed{p}\sigma^{}_\mu &=& 0,\\
\sum_\mu\sigma^{}_\mu \slashed{k}(\slashed{k}+\slashed{p})\sigma^{}_\mu &=& 2(\slashed{k}+\slashed{p})\slashed{k}. 
\end{eqnarray}
This leads after performing the trace to the following expression:
\begin{widetext}
\begin{eqnarray} 
C^{(1)}_{g,1} = \left. - g
\int \frac{d^2k}{(2\pi)^2}\int \frac{d^2q}{(2\pi)^2}~ \frac{\partial^2}{\partial p^2_i}
\frac{(z^2+q^2)(z^2+k^2)+(z^2+k^2)q\cdot p +(z^2+q^2)k\cdot p - p^2 q\cdot k + 2k\cdot p~q\cdot p}
{(z^2 + q^2)(z^2 + k^2)(z^2 + (p+q)^2)(z^2+(p+k)^2)}
\right|_{p=0}.
\end{eqnarray}
After taking derivatives with respect to the momentum $p_i$ we get
\begin{equation}
\nn
\begin{array}{ccc}
\displaystyle
\left.\frac{A^{\prime\prime}(p)}{B(p)}\right|_{p=0} = 
\frac{4 k^{}_i q^{}_i - 2q\cdot k}
{(z^2+q^2)^2(z^2+k^2)^2};
&&
\displaystyle
\left. \frac{A^\prime(p)B^\prime(p)}{B^2(p)} \right|_{p=0} = 2 
\frac{[q^{}_i(z^2+k^2)+k^{}_i(z^2+q^2)]^2}
{(z^2+q^2)^3(z^2+k^2)^3};
\\\\
\displaystyle
\left.\frac{A(p)[B^\prime(p)]^2}{B^3(p)}\right|_{p=0} = 4 
\frac{[q^{}_i(z^2+k^2)+k^{}_i(z^2+q^2)]^2}
{(z^2+q^2)^3(z^2+k^2)^3};
&&
\displaystyle
\left. \frac{A(p)B^{\prime\prime}(p)}{B^2(p)} \right|_{p=0} = 2
\frac{(z^2+k^2)+4 q^{}_ik^{}_i+(z^2+q^2)}
{(z^2+q^2)^2(z^2+k^2)^2}.
\end{array}
\end{equation}
\end{widetext}
With exception of the first expression all of them are the same as compared to the case of random gap disorder. Hence, the character of the singularity does not change, while the regular part averages to zero after performing angular integration. With the same redefinition of the cutoff as above, the result for this conductivity contribution reads
\begin{equation}
\bar\sigma^{(1)}_1 = \frac{g}{\pi^2}\left(1-\frac{\omega^4}{\Lambda^4}\right)\ln\frac{\Lambda}{\omega}. 
\end{equation}
The total contribution of all diagrams to the conductivity then becomes
\begin{equation}
\bar\sigma^{}_1 = 2\bar\sigma^{(1)}_1 + 4\bar\sigma^{(2)}_1 = 0.
\end{equation}

Now we turn our attention to BLG.  Green's function for BLG reads
\begin{equation}
\label{eq:GreenBLG}
G^{}_q(\pm iz) = \frac{(q^2_1-q^2_2)\sigma^{}_1 + 2 q^2_1 q^2_2\sigma^{}_2\mp iz}{z^2+q^4} = \frac{\slashed Q(q)\mp iz}{z^2+q^4}, 
\end{equation}
with $Q^{}_1 = q^2_1-q^2_2$ and $Q^{}_2 = 2q^{}_1q^{}_2$. For the gap disorder, the evaluation of Eq.~(\ref{eq:Correction2}) goes analogously to the MLG calculations. After reshifting of momenta, the $k$--integration can be performed separately which gives 
\begin{equation}
\int \frac{d^2k}{(2\pi)^2}~G^{}_{k}(-iz) = \frac{i}{8}.
\end{equation}
This factor can be pulled out of the trace operator and product of sigma matrices $\sigma^{}_3$, corresponding to the disorder type, gives a unity matrix. Therefore, the result of the evaluation will be the same for the scalar disorder too, while that for the vector disorder will be twice this value. After performing the trace over the remaining terms we arrive at
\begin{equation}
-\frac{g}{2} \left.\int\frac{d^2q}{(2\pi)^2}~\frac{\partial^2}{\partial p^2_1}\frac{i2z}{(z^2+q^4)^2}
\frac{z^2-q^4+2[(q^{}_1+p^{}_1)^2-q^2_2](q^{2}_1-q^{2}_2)+8(q^{}_1+p^{}_1)q^{}_1q^2_2}
{z^2+[(q^{}_1+p^{}_1)^2+q^2_2]^2}\right|_{p^{}_1=0}. 
\end{equation}
Performing the derivatives with the help of Eq.~(\ref{eq:Derivative}) we obtain exploiting the rotational invariance
\begin{equation}
\left.\frac{\partial^2}{\partial p^2_1} \frac{z^2-q^4+2[(q^{}_1+p^{}_1)^2-q^2_2](q^2_1-q^2_2)+8(q^{}_1+p^{}_1)q^{}_1q^2_2}
{z^2+[(q^{}_1+p^{}_1)^2+q^2_2]^2}\right|_{p^{}_1=0} 
= -\frac{8q^2}{z^2+q^4}.
\end{equation}
The contribution to the displacement function from this diagram then reads 
\begin{equation}
C^{(1)}_{g,2} = -g \int\frac{d^2q}{(2\pi)^2}~\frac{q^2}{(z^2+q^4)^3} = -\frac{g}{16\pi z^3} = -\frac{g}{2\pi\omega^3},
\end{equation}
for both gap and scalar disorder, and twice this value for the vector disorder. The conductivity correction for gap/scalar disorder is therefore proportional to $\omega^{-1}$:
\begin{equation}
\bar\sigma^{(1)}_2 = - \frac{g}{4\pi\omega} = - \frac{g\bar\sigma^{}_0}{8\omega},
\end{equation}
where the minimal conductivity of BLG is $\bar\sigma^{}_0 = 2/\pi$. For the vector disorder it must be multiplied by 2.

The type of disorder matters if we evaluate the contribution to the displacement function from the first diagram in Fig.~\ref{fig:F1} given  analytically in Eq.~(\ref{eq:Correction1}). In order to simplify the calculation we reshift momenta as in the MLG case discussed above. In terms of Eq.~(\ref{eq:Derivative}) 
\begin{equation}
C^{(1)}_{g,1} = -\frac{g}{2}\left.\int\frac{d^2q}{(2\pi)^2}\int\frac{d^2k}{(2\pi)^2}~\frac{\partial^2}{\partial p^2_1}\frac{A(p)}{B(p)}\right|_{p=0},
\end{equation}
we obtain for all disorder types the same denominator
\begin{equation}
B(p) = (z^2+q^4)(z^2+k^4)(z^2+(q+p)^4)(z^2+(k+p)^4),
\end{equation}
but different expressions for numerators. For scalar/gap disorder it can be written as
\begin{eqnarray}
\nn
A(p) &=& 2\left\{ 
z^4 + z^2\left[ 
Q(q)\cdot Q(q+p) + Q(k)\cdot Q(k+p) 
\right. 
\right.
\\
\nn
&\mp& \left.\left(Q(q)\cdot Q(k) + Q(q+p)\cdot Q(k+p) - 
Q(q)\cdot Q(k+p) - Q(k)\cdot Q(q+p)\right)\right]\\
\nn
&+& \left. Q(q)\cdot Q(q+p)~Q(k)\cdot Q(k+p) - Q(q)\cdot Q(k+p)~Q(k)\cdot Q(q+p) + Q(q)\cdot Q(k)~Q(q+p)\cdot Q(k+p)\right\}
\end{eqnarray}
with ($-$)/($+$) for scalar/gap disorder. Using the relations $\sum_\mu \sigma^{}_\mu \slashed Q\sigma^{}_\mu=0$ and $\sum_\mu\sigma^{}_\mu \slashed Q(q)\slashed Q(q+p)\sigma^{}_\mu = 2\slashed Q(q+p)\slashed Q(q)$, the numerator for the vector disorder becomes after performing the trace 
\begin{eqnarray}
\nn
A(p) &=& 4\left\{ 
z^4 + z^2 [Q(q)\cdot Q(q+p) + Q(k)\cdot Q(k+p)] 
\right.\\
\nn
&+&\left. Q(q)\cdot Q(q+p)~Q(k)\cdot Q(k+p) - Q(q)\cdot Q(k)~Q(q+p)\cdot Q(k+p) + Q(q)\cdot Q(k+p)~Q(k)\cdot Q(q+p) 
\right\}.
\end{eqnarray}
After performing the derivatives, taking the limit $p^{}\to0$ and carrying out integrations we obtain for scalar/gap disorder in terms of Eq.~(\ref{eq:Derivative}):
\begin{eqnarray}
\int\frac{d^2q}{(2\pi)^2} \int\frac{d^2k}{(2\pi)^2}~\left.\frac{A^{\prime\prime}(p)}{B(p)}\right|_{p=0} &=& 0, \\
-2\int\frac{d^2q}{(2\pi)^2} \int\frac{d^2k}{(2\pi)^2}~\left. \left( 
\frac{A^\prime(p)B^\prime(p)}{B^2(p)}-\frac{A(p)[B^\prime(p)]^2}{B^3(p)}
\right)\right|_{p=0} &=& \frac{1}{4\pi z^3},\\
-\int\frac{d^2q}{(2\pi)^2} \int\frac{d^2k}{(2\pi)^2}~\left. 
\frac{A(p)B^{\prime\prime}(p)}{B^2(p)}
\right|_{p=0} &=& -\frac{1}{2\pi z^3}.
\end{eqnarray}
Thefore, the contribution to the displacement function for scalar/gap disorder becomes with $z = \omega/2$
\begin{equation}
C^{(1)}_{g,1} = \frac{g}{\pi \omega^3},
\end{equation}
and the corresponding conductivity correction
\begin{equation}
\bar\sigma^{(1)}_1 = \frac{g\bar\sigma^{}_0}{4\omega}. 
\end{equation}
In the case of vector disorder we have
\begin{eqnarray}
\int\frac{d^2q}{(2\pi)^2} \int\frac{d^2k}{(2\pi)^2}~\left.\frac{A^{\prime\prime}(p)}{B(p)}\right|_{p=0} &=& 0, \\
-2\int\frac{d^2q}{(2\pi)^2} \int\frac{d^2k}{(2\pi)^2}~\left. \left( 
\frac{A^\prime(p)B^\prime(p)}{B^2(p)}-\frac{A(p)[B^\prime(p)]^2}{B^3(p)}
\right)\right|_{p=0} &=& \frac{1}{2\pi z^3},\\
-\int\frac{d^2q}{(2\pi)^2} \int\frac{d^2k}{(2\pi)^2}~\left. 
\frac{A(p)B^{\prime\prime}(p)}{B^2(p)}
\right|_{p=0} &=& -\frac{1}{\pi z^3},
\end{eqnarray}
which yields for the displacement function correction 
\begin{equation}
C^{(1)}_{g,1} = \frac{2g}{\pi \omega^3},
\end{equation}
and eventually for the conductivity correction
\begin{equation}
\bar\sigma^{(1)}_1 = \frac{g\bar\sigma^{}_0}{2\omega}. 
\end{equation}
In all cases, the total first--order conductivity correction is zero:
\begin{equation}
\bar\sigma^{(1)} = 2\bar\sigma^{(1)}_1 + 4\bar\sigma^{(1)}_2  = 0.
\end{equation}

\section{Evaluation algorithm for higher--order diagrams}

The higher--order Feynman graphs were evaluated with the help of computer algebra package Mathematica. We used the algorithm described below:
\begin{itemize}
\item Draw diagrams and distribute momentum indices in accord with momentum conservation;
\item Reshift momenta appropriately, perform the trace and introduce the polar coordinates;
\item Perform derivatives with respect to $p^{}_1$ and send $p\to0$. For higher--order diagrams it is convenient to use Eq.~(\ref{eq:Derivative});
\item Perform angular integrals. At this stage all second--order diagrams containing crossed impurity lines vanish;
\item Perform radial integrals from 0 to the upper cutoff $\Lambda$. Replace $z=\omega/2$ and redefine for MLG the cutoff via  $\Lambda\to\sqrt{\Lambda^2-\omega^2}/2$. For BLG, the limit $\Lambda\to\infty$ is carried out. Use the command FullSimplify[{\it expression}, Assumptions$\to\{\Lambda>0,~\omega>0\}$];
\item Sum up contributions from all diagrams taking correct degeneracy factors into account, Use the command FullSimplify[{\it expression}, Assumptions$\to\{\Lambda>0,~\omega>0\}$]. At this stage all divergent contributions drop out;
\item Multiply with $-\omega^2/2(g/2)^n$, where $n$ denotes the diagram order and perform the limit $\Lambda\to\infty$. Only the finite corrections survive. 
\end{itemize}

\begin{figure}[t]
\includegraphics[width=10cm]{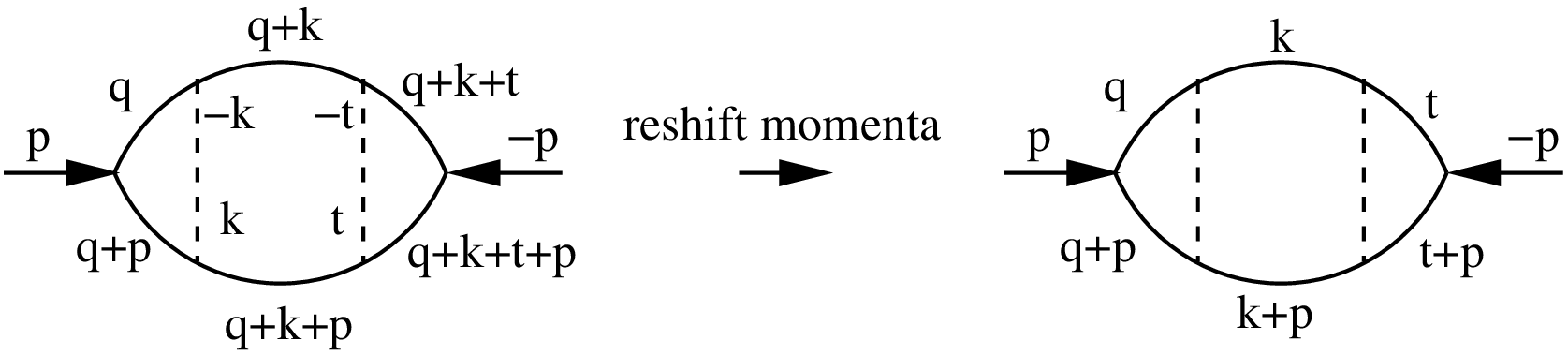}  
\caption{Momentum reshifting of the second-order ladder diagram.}
\label{fig:F2}
\end{figure}

{\bf Example 1:} As a first example we consider the second--order ladder diagram for MLG with random scalar disorder shown in Fig.~\ref{fig:F2}. 
\begin{enumerate}
\item First we define propagators. The Mathematica code reads:
\begin{itemize}
 \item Define Pauli matrices:
$$
\nn
\begin{array}{lcr}
\sigma^{}_0 &=& \{\{1, 0\}, \{0, 1\}\}\\
\sigma^{}_1 &=& \{\{0, 1\}, \{1, 0\}\}\\
\sigma^{}_2 &=& \{\{0,-{\rm I}\}, \{{\rm I}, 0\}\}\\
\end{array}
$$
\item Define propagators for positive and negative frequencies:
$$
\nn
\begin{array}{lcr}
{\rm GreenPl} &=& (-{\rm I}~z \sigma_0 + {\rm q1}~\sigma_1 + {\rm q2}~\sigma_2)/(z^2 +{\rm q1^2 + q2^2}) \\
{\rm GreenMi} &=& (\;\;\;{\rm I}~z \sigma_0 + {\rm q1}~\sigma_1 + {\rm q2}~\sigma_2)/(z^2 +{\rm q1^2 + q2^2})
\end{array}
$$
\end{itemize}

\item Now we draw the diagram and put momentum indices as required by momentum conservation. This is displayed by the left graph in Fig.~\ref{fig:F2}. Next we reshift momenta in two steps: 1) $t\to t-q-k$ and 2) $k\to k-q$. The result can be seen on the right graph in Fig.~\ref{fig:F2}. The Mathematica code which implements these steps reads: 
$$
\nn
\begin{array}{rcl}
{\rm B1} &=& {\rm GreenMi}\\
{\rm B2} &=& {\rm GreenPl} \diagup\ldotp \{{\rm q1 \to q1 + p1,~ q2 \to q2 + p2}\} \\
{\rm B3} &=& {\rm GreenPl} \diagup\ldotp \{{\rm q1 \to q1 + k1 + p1,~ q2 \to q2 + k2 + p2}\} \\
{\rm B4} &=& {\rm GreenPl} \diagup\ldotp \{{\rm q1 \to q1 + k1 + t1 + p1,~ q2 \to q2 + k2 + t2 + p2}\} \\
{\rm B5} &=& {\rm GreenMi} \diagup\ldotp \{{\rm q1 \to q1 + k1 + t1,~ q2 \to q2 + k2 + t2}\} \\
{\rm B6} &=& {\rm GreenMi} \diagup\ldotp \{{\rm q1 \to q1 + k1,~ q2 \to q2 + k2}\}
\end{array}
$$
Perform the trace, reshift momenta as discussed above, put $\rm p2\to 0$ and introduce polar coordinates
$$
\nn
\begin{array}{rcl}
{\rm A1} &=& {\rm Simplify}[(({\rm (Dot[B1, B2, B3, B4, B5, B6][[1, 1]] + 
       Dot[B1, B2, B3, B4, B5, B6][[2, 2]])/(2~Pi)^6}\\
&& \diagup\ldotp \{\rm t1 \to t1 - k1 - q1,~ t2 \to t2 - k2 - q2\})\\
&& \diagup\ldotp \{\rm k1 \to k1 - q1,~ k2 \to k2 - q2 \})\\
&& \diagup\ldotp  {\rm \{p2 \to 0,~ q1 \to q~ Cos[x],~ q2 \to q~ Sin[x],~
k1 \to k~Cos[y],~ k2 \to k~Sin[y],~ t1 \to t~Cos[u],~ t2 \to t~Sin[u]\}}]
\
\end{array}
$$

\item Perform derivatives with respect to p1, put $\rm p1\to0$ and perform angular integrations
$$
\nn
\begin{array}{rcl}
{\rm  A2} = {\rm Integrate[Integrate[Integrate[Simplify[D[A1, p1, p1] \diagup\ldotp p1 \to 0], \{x,~0,~ 2~ Pi\}], \{y,~0,~2~ Pi\}], \{u,~0,~2~ Pi\}]}
\end{array}
$$

\item Perform radial integrations from 0 to $\Lambda$:
$$
\nn
\begin{array}{rcl}
{\rm A3} &=& {\rm (Integrate[t~ A2,~ t] \diagup\ldotp  t \to \Lambda) - (Integrate[t~ A2,~ t] \diagup\ldotp t \to 0) } \\ 
{\rm A4} &=& {\rm (Integrate[k~ A3,~ k] \diagup\ldotp  k \to \Lambda) - (Integrate[k~ A3,~ k] \diagup\ldotp k \to 0) } \\
{\rm A5} &=& {\rm (Integrate[q~ A4,~ q] \diagup\ldotp  q \to \Lambda) - (Integrate[q~ A4,~ q] \diagup\ldotp q \to 0) }
\end{array}
$$

\item Simplify, multiply with $-\omega^2/2$, reshift the cutoff by $\Lambda\to\sqrt{\Lambda^2-\omega^2}/2$ and simplify again
$$
\nn
\begin{array}{rcl}
{\rm A6} &=& {\rm FullSimplify[A5, Assumptions \to {\Lambda > 0,~ z > 0}]}\\
{\rm A7} &=&
{\rm FullSimplify[-\omega^2/2 ~ (A6 \diagup\ldotp \{z\to \omega/2,~\Lambda \to Sqrt[\Lambda^2 - \omega^2]/2\}),~ Assumptions \to \{\Lambda > 0,~\omega > 0\}]}
\end{array}
$$

The result of the last operation reads
$$
\frac{\displaystyle (\Lambda^2-\omega^2)^3 + 2\Lambda^2\ln\frac{\Lambda}{\omega}\left(2(\Lambda^2-\omega^2)^2 +3(\Lambda^4-\omega^4)\ln\frac{\Lambda}{\omega}\right)}
{8\pi^3\Lambda^6},
$$
\end{enumerate}
which upon multiplication with $(g/2)^2$ and retaining only finite and divergent terms for $\Lambda\to\infty$ yields the conductivity correction 
$$\displaystyle 
\sigma^{(1)}_2 = \bar\sigma^{}_0\left( \frac{1}{8}\frac{g^2}{(2\pi)^2}+\frac{1}{2}\frac{g^2}{(2\pi)^2}\ln\frac{\Lambda}{\omega}+\frac{3}{4}\left[\frac{g}{2\pi}\ln\frac{\Lambda}{\omega}\right]^2
\right),
$$
where $\bar\sigma^{}_0=1/\pi$, as it can be seen in Table~II in the main text.

\begin{figure}[t]
\includegraphics[width=10cm]{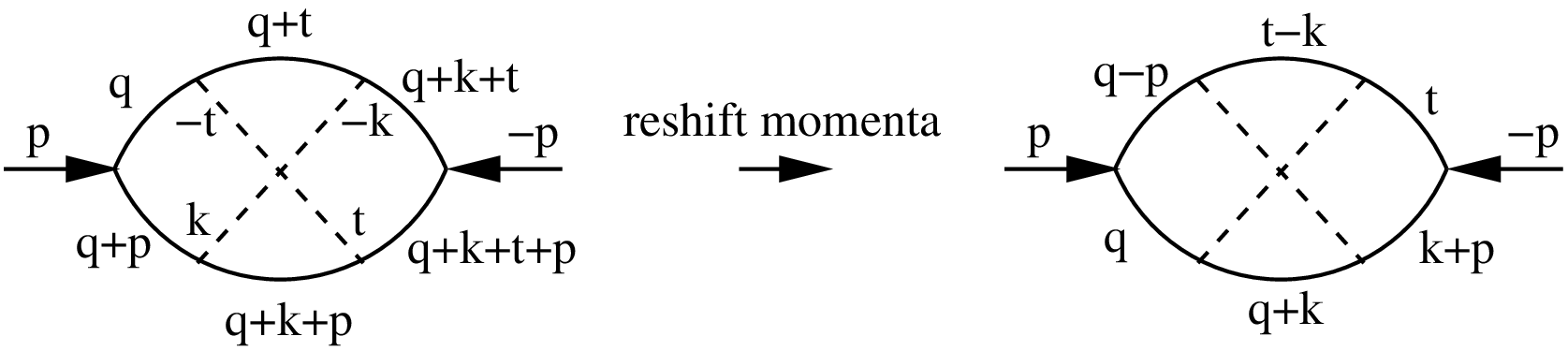}  
\caption{Momentum reshifting of the second-order X--diagram.}
\label{fig:F3}
\end{figure}

{\bf Example 2:} As another example we demonstrate vanishing of the so--called diagram with maximally crossed impurity lines shown in Fig.~\ref{fig:F3} for MLG.

\begin{enumerate}

\item First we distribute momenta in accord with the momentum conservation as shown on the left in Fig.~\ref{fig:F3} and reshift them in the following steps: $1)~ t\to t-k-q$, $2)~ q \to q-p$. The result is shown on the right in  Fig.~\ref{fig:F3}. To implement this with Mathematica we have to define the propagators:
$$
\nn
\begin{array}{rcl}
{\rm B1} &=& {\rm GreenMi}\\
{\rm B2} &=& {\rm GreenPl} \diagup\ldotp \{{\rm q1 \to q1 + p1,~ q2 \to q2 + p2}\} \\
{\rm B3} &=& {\rm GreenPl} \diagup\ldotp \{{\rm q1 \to q1 + k1 + p1,~ q2 \to q2 + k2 + p2}\} \\
{\rm B4} &=& {\rm GreenPl} \diagup\ldotp \{{\rm q1 \to q1 + k1 + t1 + p1,~ q2 \to q2 + k2 + t2 + p2}\} \\
{\rm B5} &=& {\rm GreenMi} \diagup\ldotp \{{\rm q1 \to q1 + k1 + t1,~ q2 \to q2 + k2 + t2}\} \\
{\rm B6} &=& {\rm GreenMi} \diagup\ldotp \{{\rm q1 \to q1 + t1,~ q2 \to q2 + t2}\}
\end{array}
$$
multiply them, perform the trace and reshift momenta as shown above:
$$
\begin{array}{rcl}
{\rm B7} &=& {\rm Simplify[(((Simplify[Dot[B1, B2, B3, B4, B5, B6][[1, 1]] + Dot[B1, B2, B3, B4, B5, B6][[2, 2]]])} \\
& & \diagup\ldotp {\rm \{t1 \to t1 - k1 - q1,~ t2 \to t2 - k2 - q2\})\diagup\ldotp \{q1 \to q1 - p1,~ q2 \to q2 - p2\})]}
\end{array}
$$

\item Now we put p2 to zero and introduce polar coordinates:
$$
\begin{array}{rcl}
{\rm T1} &=& {\rm Simplify[B7 \diagup\ldotp\{p2 \to 0,~ q1 \to q~ Cos[x],~ q2 \to q~ Sin[x],~ k1 \to k~ Cos[y],}\\
&& {\rm  k2 \to k~ Sin[y],~ t1 \to t~ Cos[u],~ t2 \to t~ Sin[u]\}]}
\end{array}
$$

\item In order to perform differentiations with respect to p1 we use Eq.~(\ref{eq:Derivative}). The numerator (A) and denominator (B) of T1 are separated out using following commands:
$$
\begin{array}{rcl}
{\rm A} &=& {\rm Numerator[T1]}\\
{\rm B} &=& {\rm Denominator[T1]} 
\end{array}
$$

\item Now we generate first term from Eq.~(\ref{eq:Derivative}), put $\rm p1\to0$ and perform integration over the angle y from 0 to $2\pi$:
$$
\begin{array}{rcl}
{\rm T2} &=& {\rm Simplify[(Integrate[Simplify[D[A,~p1,~p1]/B \diagup\ldotp \{p1 \to 0\}],~y] \diagup\ldotp 
     y \to 2 Pi)}\\
&-& {\rm (Integrate[Simplify[D[A,~p1,~p1]/B \diagup\ldotp \{p1 \to 0\}],~ y] \diagup\ldotp y \to 0)]}
\end{array}
$$
The result of this integration reads
$$
-\frac{\rm 4 \pi z^2 \cos[u + x]}{\rm q~ t~ (q^2 + z^2)^2 (t^2 + z^2)^2}
$$
and the next integration with respect to x or u yields zero.

\item Generate second term from Eq.~(\ref{eq:Derivative}), put $\rm p1\to0$ and perform integration over the angle y:
$$
\begin{array}{rcl}
{\rm T3} &=& {\rm Simplify[(Integrate[Simplify[D[A,~p1] D[B,~p1]/B^2 \diagup\ldotp {p1 \to 0}],~y] \diagup\ldotp y \to 2 Pi)}\\
&-& {\rm (Integrate[Simplify[D[A,~p1] D[B,~p1]/B^2 \diagup\ldotp {p1 \to 0}],~y] \diagup\ldotp y \to 0)]} 
\end{array}
$$
The result is zero.

\item Generate third term from Eq.~(\ref{eq:Derivative}), put $\rm p1\to0$ and perform integration over the angle y:
$$
\begin{array}{rcl}
{\rm T4} &=& {\rm FullSimplify[(Integrate[Simplify[(A (D[B,~p1])^2/B^3) /. {p1 \to 0}],~y] \diagup\ldotp y \to 2 Pi)}\\
&-& {\rm (Integrate[Simplify[(A (D[B,~p1])^2/B^3) /. {p1 \to 0}],~y] \diagup\ldotp y \to 0)] } 
\end{array}
$$
The result is zero.

\item Generate fourth term from Eq.~(\ref{eq:Derivative}), put $\rm p1\to0$ and perform integration over the angle y:
$$
\begin{array}{rcl}
{\rm T5} &=& {\rm Simplify[(Integrate[Simplify[(A/B^2)~ D[B,~p1,~p1] \diagup\ldotp {p1 \to 0}],~y] \diagup\ldotp y \to 2 Pi)}\\
&-& {\rm (Integrate[Simplify[(A/B^2)~ D[B,~p1,~p1] \diagup\ldotp {p1 \to 0}],~ y] \diagup\ldotp y \to 0)]}
\end{array}
$$
The result is zero.
\end{enumerate}